\documentclass[%
 reprint,
superscriptaddress,
 amsmath,amssymb,
 aps,
]{revtex4-2}

\usepackage{graphicx}
\usepackage{dcolumn}
\usepackage{bm}
\usepackage{multirow}
\usepackage[normalem]{ulem}
\usepackage{xcolor}
\usepackage{hyperref}
\usepackage{comment}

\newcommand{\mue}{\tilde{\mu}_\mathrm{e}}
\newcommand{\EF}{E_\mathrm{F0}}
\newcommand{\ET}{E_\mathrm{T0}}

\newcommand{\sech}{\mathrm{sech}}
\usepackage{xcolor}

\begin{document}

\preprint{APS/123-QED}

\title{Universal scaling of electrochemical information transfer at solid--liquid interfaces}

\author{Eikichi Kimura}
\affiliation{School of Engineering, Institute of Science Tokyo, Yokohama, Kanagawa 226-8501, Japan}

\author{Ashutosh Rathi}
\affiliation{Department of Chemistry, Graduate School of Environmental, Life, Natural Science and Technology, Okayama University, Okayama, Okayama 700-8530, Japan}

\author{Saki Tsuchiya}
\affiliation{School of Engineering, Institute of Science Tokyo, Yokohama, Kanagawa 226-8501, Japan}

\author{El Mustapha Mansouri}
\affiliation{School of Engineering, Institute of Science Tokyo, Yokohama, Kanagawa 226-8501, Japan}

\author{Sara Mandi\'c}
\affiliation{Department of Chemistry, Graduate School of Environmental, Life, Natural Science and Technology, Okayama University, Okayama, Okayama 700-8530, Japan}

\author{Kristi Kanya Saikia}
\affiliation{Department of Chemistry, Graduate School of Environmental, Life, Natural Science and Technology, Okayama University, Okayama, Okayama 700-8530, Japan}

\author{Keisuke Oshimi}
\affiliation{School of Engineering, Institute of Science Tokyo, Yokohama, Kanagawa 226-8501, Japan}

\author{Masazumi Fujiwara}
\email{masazumi@okayama-u.ac.jp}
\affiliation{Department of Chemistry, Graduate School of Environmental, Life, Natural Science and Technology, Okayama University, Okayama, Okayama 700-8530, Japan}

\author{Keigo Arai}
\email{arai.k.835f@m.isct.ac.jp}
\affiliation{School of Engineering, Institute of Science Tokyo, Yokohama, Kanagawa 226-8501, Japan}

\date{\today}

\begin{abstract}
Electrochemical potentials at solid--liquid interfaces govern diverse chemical and energy conversion processes; however, the extent to which their influence extends into the solid remains unclear. Here, we show that the maximum accessible electrochemical information is governed by a single dimensionless screening parameter $u$, defined as the ratio of the effective electrostatic separation between the interface and probe to the electrostatic propagation length in the solid. The maximum Fisher information follows a universal inverse-square scaling with this parameter. This relation identifies electrostatic screening as a fundamental constraint on information transfer across solid--liquid interfaces and provides quantitative design principles for interfacial electrochemical sensing.

\end{abstract}

\maketitle


Chemical reactivity and energy conversion in a wide range of chemical and biological systems are governed by electrochemical potentials at solid--liquid interfaces~\cite{bard_electrochemical_2022,stamenkovic_energy_2017,sies_fundamentals_2024}. These potentials are established within nanometer-scale interfacial regions, where ionic screening in the liquid, surface chemical equilibria, and the electronic structure of the solid are strongly coupled~\cite{nozik_physical_1996,memming_semiconductor_2015}. The resulting electrostatic response is screened on both sides of the interface: by the electrical double layer (EDL) in the electrolyte and by charge redistribution or depletion-region formation in the solid~\cite{zhang_band_2012,wu_understanding_2022}. This raises the fundamental question of how far interfacial electrochemical information remains detectable inside the solid and which physical parameters ultimately limit its accessibility.

Recent advances in nanoscale probes provide a means to address this question through localized electronic states whose occupancies depend on the local electrochemical environment. In particular, near-surface solid-state defects~\cite{wolfowicz_quantum_2021,rathi_engineering_2026a} provide a well-defined realization, enabling interfacial electrostatic potentials to be inferred from their charge-state response~\cite{hauf_chemical_2011,grotz_charge_2012,broadway_spatial_2018,pfender_protecting_2017,mccloskey_diamond_2022,bian_nanoscale_2021}. However, a quantitative framework for the fundamental limits governing this information transfer from the interface to near-surface probes has not been formulated.

\begin{figure}[b]
  \includegraphics[width=\columnwidth]{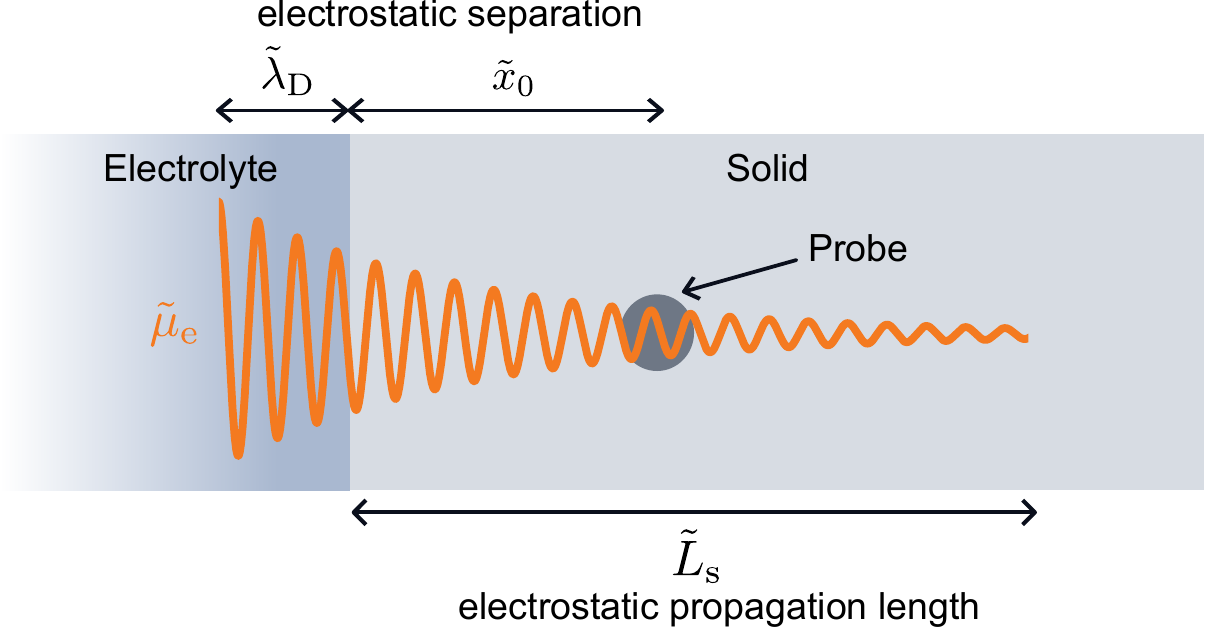}
  \caption{Encoding of electrochemical information across a solid--liquid interface. The liquid is characterized by the electrical Debye screening length $\tilde{\lambda}_\mathrm{D}$, and the near-surface probe is located at an electrical depth $\tilde{x}_0$ below the interface. Their sum, $\tilde{x}_0+\tilde{\lambda}_\mathrm{D}$, defines the total electrostatic separation between the electrochemical environment and the probe. Within the solid, electrochemical information propagates over the characteristic electrical length scale $\tilde{L}_\mathrm{s}$, referred to as the electrostatic propagation length. The wavy orange line schematically represents the flow and attenuation of electrochemical information, with the decreasing amplitude indicating attenuation.}
  \label{fig:concept}
\end{figure}

In this Letter, we demonstrate that the accessible electrochemical information is governed by a single dimensionless parameter $u$, defined as the ratio of the combined electrostatic distance set by the Debye screening length $\tilde{\lambda}_\mathrm{D}$ and near-surface probe depth $\tilde{x}_0$ to the electrostatic propagation length in the solid $\tilde{L}_\mathrm{s}$, as illustrated in Fig.~\ref{fig:concept}. The binary readout of the two-level probe carries Fisher information $\mathcal{I}(\mue)$, whose maximum is
\begin{align}
\mathcal{I}^{*} = \mathcal{I}_{0}\,\frac{1}{(1 + u)^2}, \qquad u \equiv \frac{\tilde{x}_{0}+\tilde{\lambda}_{\mathrm{D}}}{\tilde{L}_{\mathrm{s}}},
\label{eq:Fisher_main}
\end{align}
where $\mathcal{I}_{0}=1/(2k_{\mathrm{B}}T)^{2}$ is the thermally limited Fisher information. This scaling relation provides a concise framework for quantifying electrochemical information transfer across solid--liquid interfaces and guiding the design of near-surface probes for electrochemical sensing.

\begin{figure}[t]
\includegraphics[width=\columnwidth]{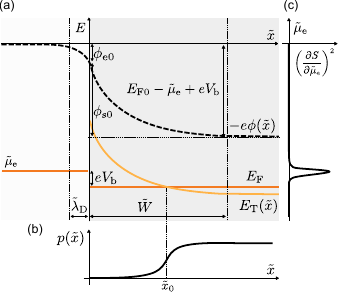}
\caption{Model of a solid--liquid interface. (a) Electrostatic energy diagram along the electrical coordinate $\tilde{x}=x/\varepsilon$. The mismatch $E_{\mathrm{F0}}-\mue+eV_{\mathrm{b}}$ is distributed between the liquid EDL, of width $\tilde{\lambda}_{\mathrm{D}}$, and the solid depletion layer, of width $\tilde{W}$, producing the potential $\phi(\tilde{x})$. A defect with charge-transition energy $\ET$ is located at depth $\tilde{x}_0$. (b) Its charge-state probability follows the Fermi--Dirac distribution and defines the signal $S$. (c) At fixed $\tilde{x}_0$, $S$ is sensitive to $\mue$ only within a limited electrochemical-potential range.}
\label{fig:modeling}
\end{figure}

We consider a minimal electrostatic model of a planar solid--liquid interface with the liquid occupying $\tilde{x}<0$ and the solid occupying $\tilde{x}>0$, as shown in Fig.~\ref{fig:modeling}. We employ the electrical length coordinate $\tilde{x}=x/\varepsilon$, where $\varepsilon=\varepsilon_{\mathrm{e}}$ for the liquid and $\varepsilon=\varepsilon_{\mathrm{s}}$ for the solid. The liquid is described within the Debye--H\"uckel approximation~\cite{bard_electrochemical_2022} under the assumption of sufficiently weak surface potential (see Sec.~S1 of the Supplemental Material). The solid is treated as a dielectric or semiconductor with Fermi level $\EF$. The electron-accumulation regime $\mue>\EF$ is excluded from the present model. The potential drop established by thermal equilibration is distributed across the EDL in the liquid and the depletion region in the solid. The corresponding electrostatic potential $\phi(\tilde{x})$ is obtained analytically by imposing continuity of the electrostatic potential and displacement field at the interface, as detailed in Sec.~S2 of the Supplemental Material. Additionally, the bias voltage $V_\mathrm{b}$ is used to shift the Fermi level and total potential drop.

To probe this interfacial electrostatic environment, we consider a near-surface two-level system embedded in a solid at an electrical depth $\tilde{x}_0$. For a solid-state defect, this corresponds to the charge transition level $\ET$, which is shifted by the local electrostatic potential according to $E_\mathrm{T}(\tilde{x}) = \ET - e\phi(\tilde{x})$ upon reaching thermal equilibrium, where $e$ is the elementary charge. At the same time, the Fermi level pinned to the electrochemical potential~\cite{chakrapani_charge_2007} is shifted by the bias voltage as $E_\mathrm{F} = \mue - eV_\mathrm{b}$, so that the occupation probability of the two-level system follows the Fermi--Dirac distribution, $p(\tilde{x})=\left\{1+\exp\left[(E_\mathrm{T}(\tilde{x})-E_\mathrm{F})/(k_\mathrm{B}T)\right]\right\}^{-1}$, where $k_\mathrm{B}$ is the Boltzmann constant and $T$ is the temperature~\cite{das_what_2020}. We define the measurement signal with this occupation probability as $S(\mue, V_\mathrm{b}) \equiv p(\tilde{x}_0)$, which corresponds to a binary readout of the system state. The Fisher information with respect to $\mue$,
\begin{align}
    \mathcal{I} = \frac{1}{S(1-S)}\,\left(\frac{\partial S}{\partial \mue}\right)^2,
\end{align}
quantifies the amount of information that can be extracted from the measurement. 

\begin{figure}[t]
  \includegraphics[width=\columnwidth]{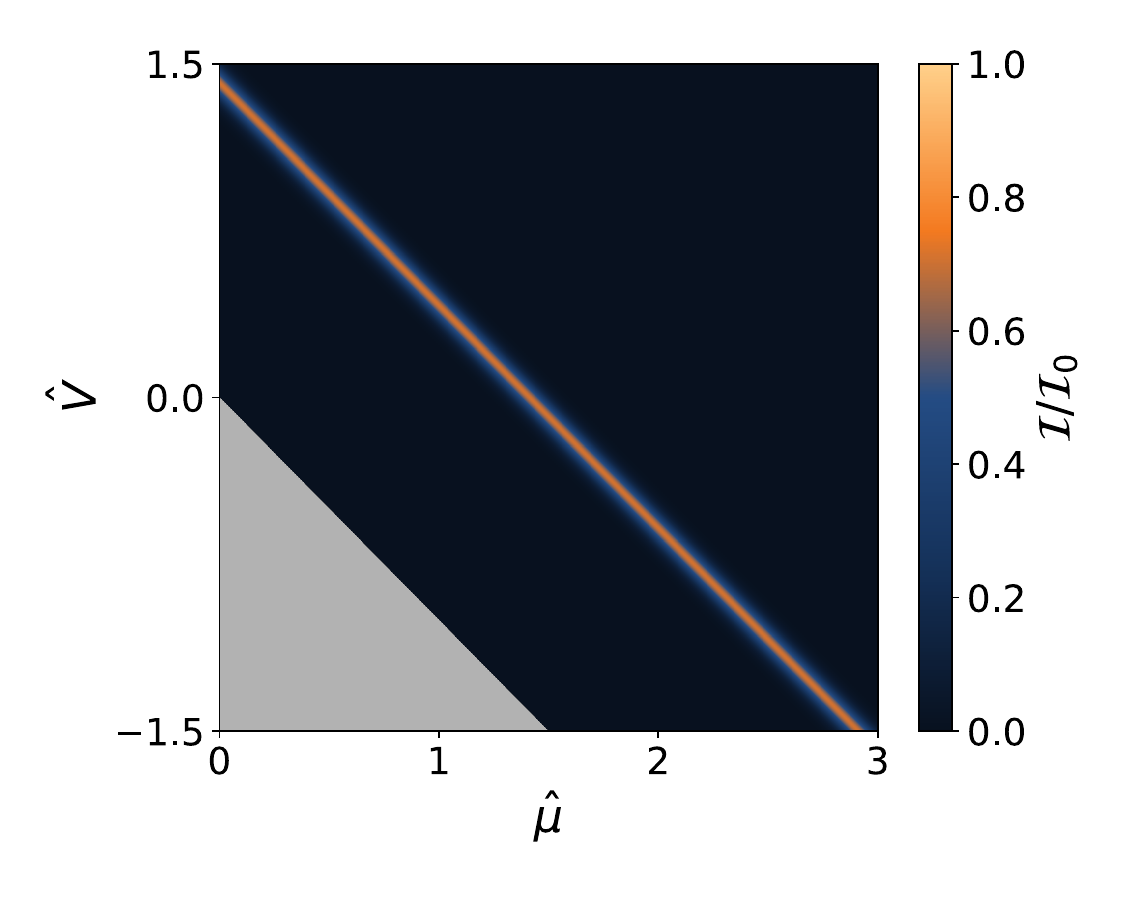}
\caption{Information landscape of electrochemical potential sensing. The information is concentrated along a narrow ridge, indicating that the electrochemical potential becomes accessible only under a restricted range of electrochemical potential and bias voltage conditions. The calculations were performed using $\tilde{\lambda}_{\mathrm{D}}=9.74~\mathrm{nm}/(80\varepsilon_0)$, $\tilde{x}_0=4.0~\mathrm{nm}/(5.6\varepsilon_0)$, and $\tilde{L}_{\mathrm{s}}=24.88~\mathrm{nm}/(5.6\varepsilon_0)$, and the gray shaded regions correspond to the accumulation regime $\EF-\mue+eV_\mathrm{b} < 0$ after applying bias voltage.}
  \label{fig:landscape}
\end{figure}

The measurement signal and Fisher information as functions of the normalized electrochemical potential $\hat{\mu}\equiv(\mue-\EF)/(\ET-\EF)$ and the normalized bias voltage $\hat{V}\equiv-eV_\mathrm{b}/(\ET-\EF)$ are shown in Fig.~\ref{fig:landscape}. The calculations were performed using $\tilde{\lambda}_{\mathrm{D}}=9.74~\mathrm{nm}/(80\varepsilon_0)$, $\tilde{x}_0=4.0~\mathrm{nm}/(5.6\varepsilon_0)$, and $\tilde{L}_{\mathrm{s}}=24.88~\mathrm{nm}/(5.6\varepsilon_0)$ (see Table~S2 of the Supplemental Material). The electrochemical potential and applied bias act linearly and oppositely on the Fermi level in the solid; increasing $\mue$ raises $E_\mathrm{F}$, whereas increasing $V_\mathrm{b}$ decreases it under our sign convention. Thus, changes in $\hat{\mu}$ can be compensated by corresponding changes in $\hat{V}$, producing the diagonal structure observed in the signal and Fisher information landscapes. Rather than being uniformly distributed over the parameter space, the Fisher information is concentrated along a narrow ridge, indicating that the interfacial electrochemical information is accessible only within a restricted range of electrostatic conditions. The shaded regions in Fig.~\ref{fig:landscape} denote the accumulation regime, defined by $\EF-\mue+eV_\mathrm{b}<0$ after application of the bias voltage, and are excluded from the present calculations (see Sec.~S1 of the Supplemental Material).

\begin{figure}[!t]
\includegraphics[width=\columnwidth]{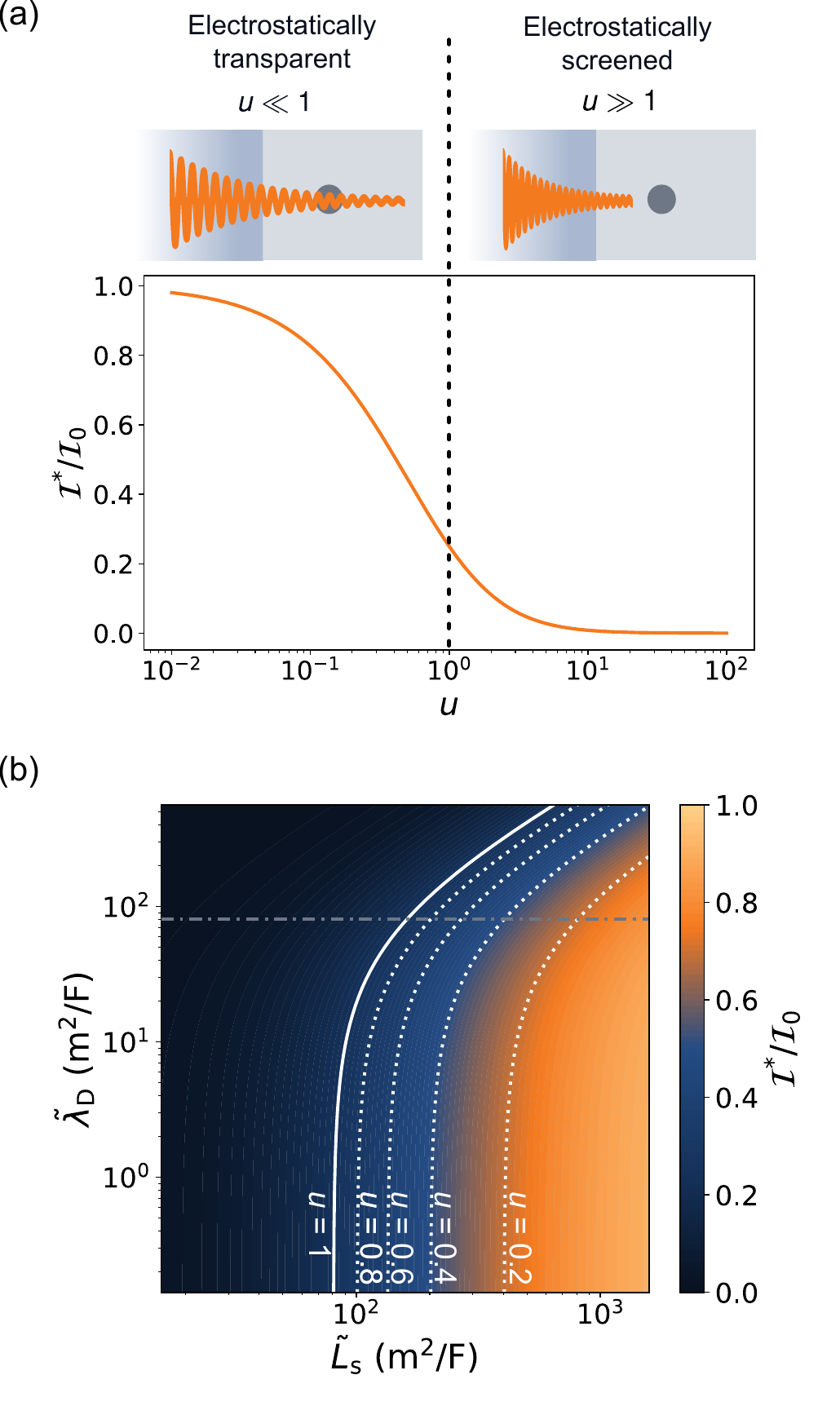}
\caption{Maximum Fisher information. (a) Universal inverse-square dependence on the screening parameter $u$, with $u\ll1$ and $u\gg1$ denoting the electrostatically transparent and strongly screened regimes, respectively. (b) Maximum information as a function of $\tilde{\lambda}_{\mathrm{D}}$ and $\tilde{L}_{\mathrm{s}}$ at fixed $\tilde{x}_0=4.0~\mathrm{nm}/(5.6\varepsilon_0)$. The contours indicate constant $u$, and the gray dash-dotted line denotes $\tilde{\lambda}_{\mathrm{D}}=\tilde{x}_0$. Information decreases with increasing $\tilde{\lambda}_{\mathrm{D}}$ and increases with $\tilde{L}_{\mathrm{s}}$.}
\label{fig:scaling}
\end{figure}

The condition for maximizing the Fisher information can be obtained analytically, yielding the optimal bias $\hat{V}^*$ as 
\begin{align}
\hat{\mu}+\hat{V}^*=\frac{(\tilde{x}_0+\tilde{L}_\mathrm{s}+\tilde{\lambda}_\mathrm{D})^2-\tilde{\lambda}_\mathrm{D}^{2}}{\tilde{L}_\mathrm{s}^{2}},
\end{align}
whose expression reveals that the optimal measurement condition is not determined by the electrochemical potential alone, but by the electrostatic geometry of the system (see Sec.~S4 of the Supplemental Material). This analytical condition agrees with direct numerical optimization to within 0.1\% over the parameter range considered here. In particular, the optimal bias voltage is determined by three electrical length scales, $\tilde{\lambda}_\mathrm{D}$, $\tilde{x}_0$, and $\tilde{L}_\mathrm{s}$. By tuning the bias to its optimum, the Fisher information reduces to Eq.~\eqref{eq:Fisher_main}. The parameter $u$ compares the effective electrostatic separation between the liquid environment and the probe with the propagation length in the solid. The regime $u \ll 1$ corresponds to the electrostatically transparent regime, whereas $u \gg 1$ corresponds to the electrostatically screened regime, as shown in Fig.~\ref{fig:scaling}(a). Therefore, maximizing the accessible Fisher information reduces to minimizing $u$.

Figure~\ref{fig:scaling}(b) shows the optimized Fisher information as a function of $\tilde{\lambda}_{\mathrm{D}}$ and $\tilde{L}_{\mathrm{s}}$ at a fixed probe depth $\tilde{x}_{0}=4~\mathrm{nm}/(5.6\varepsilon_0)$. When $\tilde{\lambda}_{\mathrm{D}}<\tilde{x}_{0}$, the electrostatic distance $\tilde{x}_{0}+\tilde{\lambda}_{\mathrm{D}}$ is dominated by the probe depth, and the Fisher information depends only weakly on $\tilde{\lambda}_{\mathrm{D}}$; once $\tilde{\lambda}_{\mathrm{D}}>\tilde{x}_{0}$, it decreases with increasing $\tilde{\lambda}_\mathrm{D}$. Similarly, increasing $\tilde{L}_{\mathrm{s}}$ enhances the accessible information regarding the interfacial electrochemical potential. Thus, $u$ can be reduced using a shallow probe and a liquid with high ionic strength or by increasing $\tilde{L}_{\mathrm{s}}$. Because $\tilde{L}_{\mathrm{s}}\propto\sqrt{(\EF-\ET)/N_{\mathrm{D}}^{+}}$, a low ionized dopant density $N_\mathrm{D}^+$ and large separation between the Fermi and transition levels provide favorable design conditions for near-surface sensing. These relationships provide clear design guidelines for improving the transfer of electrochemical information to near-surface two-level probes.

\begin{table*}[t]
\centering
\caption{Reference and fitted ionized-donor densities, fitted screening parameter $u_\mathrm{fit}$, and corresponding normalized maximum Fisher information $\mathcal{I}^*/\mathcal{I}_0$. The symmetric uncertainties are rounded 95\% confidence intervals obtained from the fits.}
\label{tab:u_Iopt}
\begin{tabular}{lcccc}
\hline\hline
Study & Reference $N_\mathrm{D}$ range ($10^{24}~\mathrm{m}^{-3}$) & $N_\mathrm{D, fit}^+$ ($10^{24}~\mathrm{m}^{-3}$) & $u_\mathrm{fit}$ & $\mathcal{I}^{*}/\mathcal{I}_0$ \\
\hline
Sow \textit{et al.}~\cite{sow_highthroughput_2020} & $8.80$--$70.4$ & $16.1 \pm 1.4$ & $0.414 \pm 0.018$ & 0.501 \\
Karaveli \textit{et al.}~\cite{karaveli_modulation_2016} & $8.80$--$70.4$ & $41.7 \pm 1.1$ & $0.237 \pm 0.003$ & 0.654 \\
Grotz \textit{et al.}~\cite{grotz_charge_2012} & $14.3$--$42.9$ & $1.54 \pm 0.04$ & $0.199 \pm 0.003$ & 0.696 \\
\hline\hline
\end{tabular}
\end{table*}

The theoretical model is examined against previous experiments using near-surface solid-state defects as probes, where the defect charge state responds to the local electrostatic environment. Sow \textit{et al.}~\cite{sow_highthroughput_2020} measured the charge-state response with varying electrochemical potential $\mue$, whereas Karaveli \textit{et al.}~\cite{karaveli_modulation_2016} and Grotz \textit{et al.}~\cite{grotz_charge_2012} investigated it under applied voltage. We fitted these datasets using the present electrostatic model, with $N_\mathrm{D}^{+}$ as the fitting parameter (see Sec.~S5 and Table~S1 of the Supplemental Material). The resulting $N_{\mathrm{D,fit}}^+$, screening parameter $u_{\mathrm{fit}}$, and corresponding normalized optimal Fisher information $\mathcal{I}^*/\mathcal{I}_0$ are summarized in Table~\ref{tab:u_Iopt}, together with the reference nitrogen density $N_\mathrm{D}$ ranges for comparison. For all three datasets, $N_\mathrm{D,fit}^+$ lies within or below the corresponding reference $N_\mathrm{D}$ ranges. This is consistent with the interpretation of $N_{\mathrm{D, fit}}^+$ as an effective ionized-donor density, whereas the reference $N_\mathrm{D}$ ranges represent estimates of the total or implanted nitrogen concentration. Direct quantitative comparison with experiments may additionally be affected by non-unity readout fidelity, incomplete transfer of the applied voltage to the solid--liquid interface, and uncertainty in determining $N_\mathrm{D}^+$. These experimental factors do not alter the underlying scaling framework, but determine how closely its predicted information limit can be approached.

Charge-state control in electrostatic environments extends beyond the defect systems considered above, having been demonstrated for group-IV vacancy centers in diamond~\cite{luhmann_chargestate_2020,zhang_neutral_2023,ilkhani_fermi_2026,pattinson_solution_2026}, as well as for defects in other solid-state platforms~\cite{delascasas_stark_2017,widmann_electrical_2019,yu_electrical_2022,white_electrical_2022,kim_electrostatic_2024,gao_interfacedriven_2026}. Although motivated by near-surface solid-state defects, the framework can, in principle, be extended to localized electronic states whose occupancies respond to the local electrostatic environment~\cite{darling_defectstate_1991}.

The present framework demonstrates a universal scaling law in which the maximum information transferable across a solid--liquid interface is governed by the single dimensionless screening parameter $u$, representing the ratio of the effective electrostatic separation to the electrostatic propagation length in the solid. The resulting Fisher information sets the Cram\'er--Rao bound, $\sigma_{\mue} \geq 1/\sqrt{\mathcal{I}(\mue)}$, providing a fundamental limit on the precision with which the interfacial electrochemical potential can be inferred~\cite{fisher_theory_1925,cover_elements_2006}. Beyond this fundamental limit, the scaling relation provides practical guidance for selecting probe depth, electrolyte screening, and ionized-donor density to improve sensitivity to the interfacial electrochemical potential. These results establish a general framework for quantifying and optimizing electrochemical information transfer from solid--liquid interfaces to localized electronic probes.

This study was supported by the JST K Program (grant number JPMJKP24F3) and by JSPS KAKENHI (grant numbers JP24H00406 and JP25K17361). E.K. and K.K.S. acknowledge support from JST SPRING under grant numbers JPMJSP2180 and JPMJSP2126, respectively. E.K. developed the analytical theory and derived the universal scaling relations. A.R., S.T., E.M.M., S.M., K.K.S., and K.O. contributed to the electrostatic framework and modeling. M.F. and K.A. conceived the project, established its conceptual framework, and supervised the project. All the authors contributed to the discussion and writing of the manuscript. The data supporting the findings of this study are publicly available~\cite{zenodo}. Additional derivations, model details, and literature-fitting
procedures are provided in the Supplemental Material~\cite{supplemental}.

\bibliography{apssamp}

\clearpage
\onecolumngrid

\setcounter{page}{1}
\setcounter{section}{0}
\setcounter{equation}{0}
\setcounter{figure}{0}
\setcounter{table}{0}

\renewcommand{\thesection}{S\arabic{section}}
\renewcommand{\theequation}{S\arabic{equation}}
\renewcommand{\thefigure}{S\arabic{figure}}
\renewcommand{\thetable}{S\arabic{table}}

\begin{center}
{\LARGE\bfseries Supplemental Material for}\\[0.6em]
{\LARGE\bfseries Universal scaling of electrochemical information transfer\\at solid--liquid interfaces}
\end{center}

\section{Validity of the model}

This section specifies the assumptions that delimit the validity of the electrostatic model introduced in the main text. The liquid is described by the Debye--H\"uckel approximation and the solid by the depletion approximation, with continuity of electrostatic potential and displacement field imposed at the planar interface.

The analytical framework is intended as a minimal equilibrium electrostatic model, and its quantitative validity is restricted by the following assumptions.

\textit{(i) Thermal equilibrium and quasistatic response.} The electron distribution in the solid and the electrochemical state of the liquid are assumed to equilibrate on a timescale much shorter than the measurement time. The probe occupancy can therefore be described by a Fermi--Dirac distribution with a Fermi level. The model does not describe strongly driven interfaces, transient charging, photoinduced nonequilibrium carrier populations, or charge-transfer kinetics that remain out of equilibrium during readout.

\textit{(ii) Debye--H\"uckel description of the electrolyte.} The liquid is treated using the linearized Poisson--Boltzmann equation~\cite{bard_electrochemical_2022,schmickler_interfacial_2010}. The corresponding Debye screening length is $\lambda_\mathrm{D}=\kappa^{-1}=\sqrt{\varepsilon_\mathrm{e}k_\mathrm{B}T/(2e^2N_\mathrm{A}I)}$, where $I$ is the ionic strength. The linearization requires $|e\phi_\mathrm{e}|/(k_\mathrm{B}T)\ll1$. Accordingly, the present framework is reliable for weak interfacial potentials in sufficiently dilute electrolytes. At larger interfacial potentials or in concentrated electrolytes, the nonlinear Poisson--Boltzmann equation or a more detailed double-layer model is preferred.

\textit{(iii) One-dimensional planar geometry.} The interface is assumed to be laterally uniform. This approximation is appropriate when the lateral feature size and radius of curvature are much larger than the relevant normal length scales, including $\lambda_D$, the depletion width $W$, and the probe depth $x_0$. For nanoparticles or patterned interfaces with comparable dimensions, curvature and additional boundary conditions are preferred.

\textit{(iv) Depletion approximation in the solid.} Within the depletion region, the mobile-carrier density is assumed negligible relative to the ionized-dopant density, so that the space-charge density can be approximated as spatially uniform~\cite{memming_semiconductor_2015,sze_physics_2007}. The present derivation is restricted to the depletion regime $\EF-\mue+eV_\mathrm{b}>0$ and does not describe electron accumulation or strong inversion. It also assumes that the depletion width is smaller than the solid dimension so that the solid can be treated as semi-infinite.

\section{Analytical modeling of the electrostatic potential}
Under the Debye--H\"uckel approximation in the liquid and the depletion approximation in the solid, the electrostatic potential takes the form
\begin{align}
  \phi_\mathrm{s}(x) &= \frac{\phi_{\mathrm{s0}}}{W^2}(x-W)^2 + \frac{\EF - \mue + eV_\mathrm{b}}{e}, \label{eq:phi_s} \\
  \phi_\mathrm{e}(x) &= \phi_{\mathrm{e0}} e^{\kappa x},\label{eq:phi_e}
\end{align}
where $W$ is the depletion width characterized by $\phi_{\mathrm{s0}} = -(eN_\mathrm{D}^+ W^2)/(2\varepsilon_\mathrm{s})$, $\phi_\mathrm{e}(x)$ for $x<0$, and $\phi_\mathrm{s}(x)$ for $x>0$. The exponential form in the liquid describes screened decay away from the interface, whereas the quadratic form in the solid follows from the spatially uniform ionized-dopant charge density assumed within the depletion region. The boundary conditions at $x=0$ are
\begin{align}
  \phi_\mathrm{s}(+0) &= \phi_\mathrm{e}(-0), \label{eq:phi_cont}\\
  \varepsilon_\mathrm{s} \left.\frac{d\phi_\mathrm{s}}{dx}\right|_{+0} &= \varepsilon_\mathrm{e} \left.\frac{d\phi_\mathrm{e}}{dx}\right|_{-0}.\label{eq:E_cont}
\end{align}
The first boundary condition fixes a common interfacial potential, whereas the second follows from the absence of an explicit sheet charge at the interface. Equations \eqref{eq:phi_s}--\eqref{eq:E_cont} yield the following depletion width:
\begin{align}
  W(\mue; V_\mathrm{b}) = -\frac{\varepsilon_\mathrm{s}}{\varepsilon_\mathrm{e} \kappa} + \frac{\varepsilon_\mathrm{s}}{e N_\mathrm{D}^+}\sqrt{\left(\frac{e N_\mathrm{D}^+}{\varepsilon_\mathrm{e} \kappa}\right)^2 + \frac{2 N_\mathrm{D}^+}{\varepsilon_\mathrm{s}}(\EF - \mue + e V_\mathrm{b})}. \label{eq:W_old}
\end{align}
Here, we introduce electrical lengths obtained by rescaling the geometric lengths by the corresponding dielectric constant, together with normalized electrochemical potential and bias voltage:
\begin{align}
    \tilde{W} &= \frac{W}{\varepsilon_\mathrm{s}}, \\
    \tilde{\lambda}_\mathrm{D} &= \frac{\lambda_\mathrm{D}}{\varepsilon_\mathrm{e}}, \\
    \tilde{x}_0 &= \frac{x_0}{\varepsilon_\mathrm{s}}, \\
    \tilde{L}_\mathrm{s} &= \frac{1}{\varepsilon_\mathrm{s}}\sqrt{\frac{2\varepsilon_\mathrm{s}}{e^2N_\mathrm{D}^+}(\EF - \ET)}, \\
    \hat{\mu} &= \frac{\mue - \EF}{\ET - \EF}, \\
    \hat{V} &= \frac{-e V_\mathrm{b}}{\ET - \EF}.
\end{align}
With these definitions, the electrical depletion width in Eq.~\eqref{eq:W_old} can then be written as
\begin{align}
  \label{eq:W}
  \tilde{W} = -\tilde{\lambda}_\mathrm{D} + \sqrt{\tilde{\lambda}_\mathrm{D}^2 + \tilde{L}_\mathrm{s}^2 (\hat{\mu} + \hat{V})}.
\end{align}

\section{Two-level measurement model}
The local transition energy at probe depth $x_0$ is
\begin{align}
    E_\mathrm{T}(x) = \ET - e\phi_\mathrm{s}(x).
\end{align}

At thermal equilibrium, the occupation probability is determined by the energy separation between the local transition level and the Fermi level. We describe this conversion using the Fermi--Dirac distribution~\cite{das_what_2020}:
\begin{align}
  p(x) = \left[1 + \exp\!\left(\cfrac{E_\mathrm{T}(x) - E_\mathrm{F}}{k_\mathrm{B} T}\right)\right]^{-1}.
\end{align}
For the ideal binary readout considered here, the measured signal is identified directly with this occupation probability:
\begin{align}
    S \equiv p(x_0),
\end{align}
Substituting the electrostatic solution from Sec.~S2 into the Fermi--Dirac expression gives the explicit dependence on the normalized electrochemical potential and bias voltage, which is explicitly written as
\begin{align}
S(\hat{\mu},\hat{V})=\left[1+\exp\!\left(\frac{1}{\hat{\theta}}\left\{1-\frac{[\tilde{x}_0-\tilde{W}(\hat{\mu},\hat{V})]^2}{\tilde{L}_\mathrm{s}^{\,2}}\right\}\right)\right]^{-1},
\end{align}
where $\hat{\theta} = (k_\mathrm{B}T)/(\ET - \EF)$ denotes the normalized thermal energy. The measurement signal as a function of $(\hat{\mu}, \hat{V})$ is illustrated in Fig.~\ref{fig:maximum}(a). Using the binary-readout definition given in the main text, the Fisher information becomes
\begin{align}
  \mathcal{I}(\mue) = \frac{1}{S(1-S)} \left(\frac{\partial S}{\partial \mue}\right)^2 = \mathcal{I}_0\,\frac{(\tilde{W} - \tilde{x}_0)^2}{(\tilde{W} + \tilde{\lambda}_\mathrm{D})^2}\,\sech^2\left(\frac{\eta}{2}\right), \label{eq:Fisher}
\end{align}
where $\mathcal{I}_0=(2k_\mathrm{B}T)^{-2}$ is the thermally limited information scale and $\eta$ denotes the argument of the Fermi--Dirac distribution. The Fisher information as a function of $(\hat{\mu}, \hat{V})$ is illustrated in Fig.~\ref{fig:maximum}(b).

\section{Optimal measurement condition}
\begin{figure}[t]
  \includegraphics[width=\textwidth]{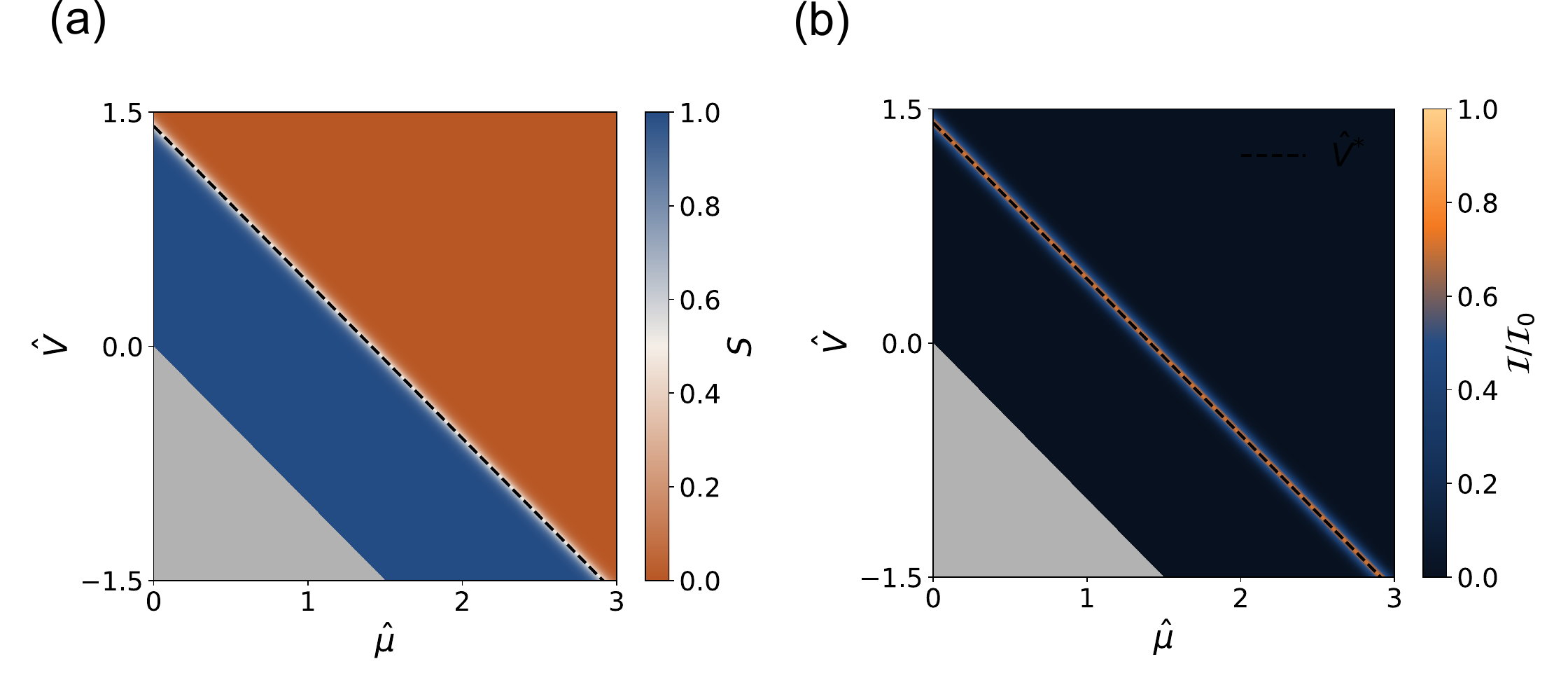}
\caption{Information landscape of electrochemical potential sensing with optimal dotted line. (a) Two-level readout signal $S$ as a function of the normalized electrochemical potential $\hat{\mu}$ and normalized bias voltage $\hat{V}$. The gray shaded region corresponds to the electron-accumulation layer, $\EF-\mue+eV_\mathrm{b} < 0$, where no depletion layer is formed. The black dotted line indicates $\eta=0$. (b) Fisher information $\mathcal{I}$ as a function of $\hat{\mu}$ and $\hat{V}$. The information is concentrated along a narrow ridge, indicating that the electrochemical potential becomes accessible only under a restricted electrochemical potential and bias voltage conditions. The black dotted line indicates $\eta=0$ and overlaps with the ridge, confirming that this analytical condition closely follows the true maximum.}
  \label{fig:maximum}
\end{figure}

The Fisher information is largest when the two-level system operates near the midpoint of its thermally broadened transition, $\eta=0$. Figure~\ref{fig:maximum} shows that this condition follows the ridge of the full Fisher-information landscape over the relevant parameter range. Numerical optimization of the Fisher information confirms that the Fisher information evaluated under the $\eta=0$ condition agrees with the exact numerical maximum to within 0.1\% over the parameter range considered here. We therefore use $\eta=0$ as the analytical optimum condition, which gives

\begin{align}
1-\frac{(\tilde{x}_0-\tilde{W}^*)^2}{\tilde{L}_\mathrm{s}^{\text{}2}}=0.
\label{eq:optimal_condition}
\end{align}

The corresponding optimal electrical depletion depth is
\begin{align}
\tilde{W}^*=\tilde{x}_0+\tilde{L}_\mathrm{s}.
\label{eq:W_opt}
\end{align}

The depletion width is controlled through the applied bias. Substituting Eq.~\eqref{eq:W_opt} into Eq.~\eqref{eq:W} therefore gives the bias condition required to place the probe at the optimal point:

\begin{align}
\hat{\mu}+\hat{V}^*=\frac{(\tilde{x}_0+\tilde{L}_\mathrm{s}+\tilde{\lambda}_\mathrm{D})^2-\tilde{\lambda}_\mathrm{D}^{2}}{\tilde{L}_\mathrm{s}^{2}}.
\end{align}

This expression shows how the operating bias compensates changes in liquid screening and solid electrostatic propagation. Its role is to tune the two-level system to the charge-transition midpoint. After this optimization, the remaining information loss is determined only by the electrostatic length-scale ratio derived below.

This optimal bias voltage yields the optimal Fisher information of
\begin{align}
\mathcal{I}^*=\mathcal{I}_0\frac{\tilde{L}_\mathrm{s}^{2}}{(\tilde{L}_\mathrm{s}+\tilde{x}_0+\tilde{\lambda}_\mathrm{D})^2},
\end{align}
which can be expressed by defining a single dimensionless parameter $u$ as
\begin{align}
\mathcal{I}^*=\mathcal{I}_0\frac{1}{(1+u)^2}, \quad u\equiv\frac{\tilde{x}_0+\tilde{\lambda}_\mathrm{D}}{\tilde{L}_\mathrm{s}}.\label{eq:Iopt_universal}
\end{align}
Equation~\eqref{eq:Iopt_universal} shows that all microscopic quantities enter the optimized Fisher information only through the dimensionless ratio $u$. Consequently, systems sharing the same $u$ exhibit identical optimized Fisher information regardless of the individual values of $\lambda_\mathrm{D}$, $x_0$, or $L_\mathrm{s}$. This collapse is the origin of the universal inverse-square scaling discussed in the main text.

\section{Literature fitting}
We fitted previously reported charge-state modulation data to test whether experimentally observed responses are consistent with the electrostatic scaling derived above. We selected the datasets of Sow \textit{et al.}~\cite{sow_highthroughput_2020}, Karaveli \textit{et al.}~\cite{karaveli_modulation_2016}, and Grotz \textit{et al.}~\cite{grotz_charge_2012} according to two criteria. First, the measurements were performed for defect centers in a solid directly interfaced with an electrolyte. Second, the measured charge-state-dependent fluorescence approaches saturation at both ends of the reported pH or voltage range. The two plateaus provide experimentally defined lower and upper levels, allowing each response to be normalized to the range \([0,1]\) without introducing an arbitrary amplitude scale.

\subsection{Parameters in the liquid}
The electron electrochemical potential at the interface was determined by assuming an equilibrium with the \(\mathrm{O_2/H_2O}\) redox couple. For the redox reaction
\begin{align}
\mathrm{O_2}+4\mathrm{H^+}+4e^-\rightleftharpoons2\mathrm{H_2O},
\end{align}
the electron electrochemical potential \(\mue\) at the interface is given by the Nernst equation:
\begin{align}
\mue=-e\left[E_{\mathrm{SHE}}^{\mathrm{abs}}+E^\circ_{\mathrm{O_2/H_2O}}-\frac{k_\mathrm{B}T}{4e}\ln\left(\frac{a_{\mathrm{H_2O}}^2}{p_{\mathrm{O_2}}a_{\mathrm{H^+}}^4}\right)\right],
\end{align}
where \(E_{\mathrm{SHE}}^{\mathrm{abs}}=4.44\ \mathrm{V}\) is the absolute potential of the standard hydrogen electrode (SHE) relative to the vacuum level, \(E^\circ_{\mathrm{O_2/H_2O}}=1.229\ \mathrm{V}\) versus SHE is the standard reduction potential of the \(\mathrm{O_2/H_2O}\) redox couple, \(a_{\mathrm{H_2O}}=1\) is the activity of water, \(p_{\mathrm{O_2}}=0.21\) is the oxygen partial pressure normalized by the standard pressure, and \(a_{\mathrm{H^+}}\) is the hydrogen ion activity. Under the ideal dilute-solution approximation, we use \(a_{\mathrm{H^+}}\simeq[\mathrm{H^+}]/c^\circ\), where \(c^\circ=1\ \mathrm{mol\,L^{-1}}\). The hydrogen ion concentration and ionic strength were selected according to the solution conditions reported in each study. The values used in these calculations are listed in Table~\ref{tab:fitting_conditions}.

\subsection{Parameters in the solid}

The surface termination was assigned according to each experiment. The acid-cleaned nanodiamonds used by Sow \textit{et al.} were treated as oxygen-terminated because the authors attribute their pH response primarily to deprotonation of oxygen-containing surface groups, particularly COOH groups~\cite{sow_highthroughput_2020}. The electron affinity was therefore set to $\chi_{\mathrm{O}}=+1.7\ \mathrm{eV}$, the experimentally reported value for chemically oxidized diamond~\cite{maier_electron_2001}. The hydrogenated nanodiamonds studied by Karaveli \textit{et al.} and the diamond surface studied by Grotz \textit{et al.} were treated as hydrogen-terminated, for which $\chi_{\mathrm{H}}=-1.1\ \mathrm{eV}$ was used~\cite{garrido_diamond_2008,takeuchi_direct_2005,tsugawa_schottky_2010}.

For both terminations, the Fermi level was placed $1.7\ \mathrm{eV}$ below the conduction-band edge~\cite{collins_fermi_2002}. Thus, $E_\mathrm{F0}=-3.40\ \mathrm{eV}$ for the oxygen-terminated Sow dataset and $E_\mathrm{F0}=-0.60\ \mathrm{eV}$ for the hydrogen-terminated Karaveli and Grotz datasets. Using a diamond band gap of $5.5\ \mathrm{eV}$, the oxygen-terminated band edges for Sow \textit{et al.} are $E_\mathrm{C}=-1.70\ \mathrm{eV}$ and $E_\mathrm{V}=-7.20\ \mathrm{eV}$; accordingly, the $\mathrm{NV}^{0/-}$ transition energy was set to $E_\mathrm{T}^{0/-}=-4.40\ \mathrm{eV}$~\cite{weber_quantum_2010}. For the hydrogen-terminated Karaveli and Grotz datasets, the $\mathrm{NV}^{+/0}$ transition energy was set to $E_\mathrm{T}^{+/0}=-3.75\ \mathrm{eV}$~\cite{grotz_charge_2012}.

\subsection{Fitting method}
We fitted the effective ionized-donor concentration \(N_\mathrm{D, fit}^+\) to the charge-state modulation reported in the literature using the present model. The fitted concentration was then used to calculate \(u_\mathrm{fit}\), which was compared to the reference range \(u_\mathrm{exp}\) inferred from the corresponding experimental conditions.

\begin{figure}[t]
  \includegraphics[width=\textwidth]{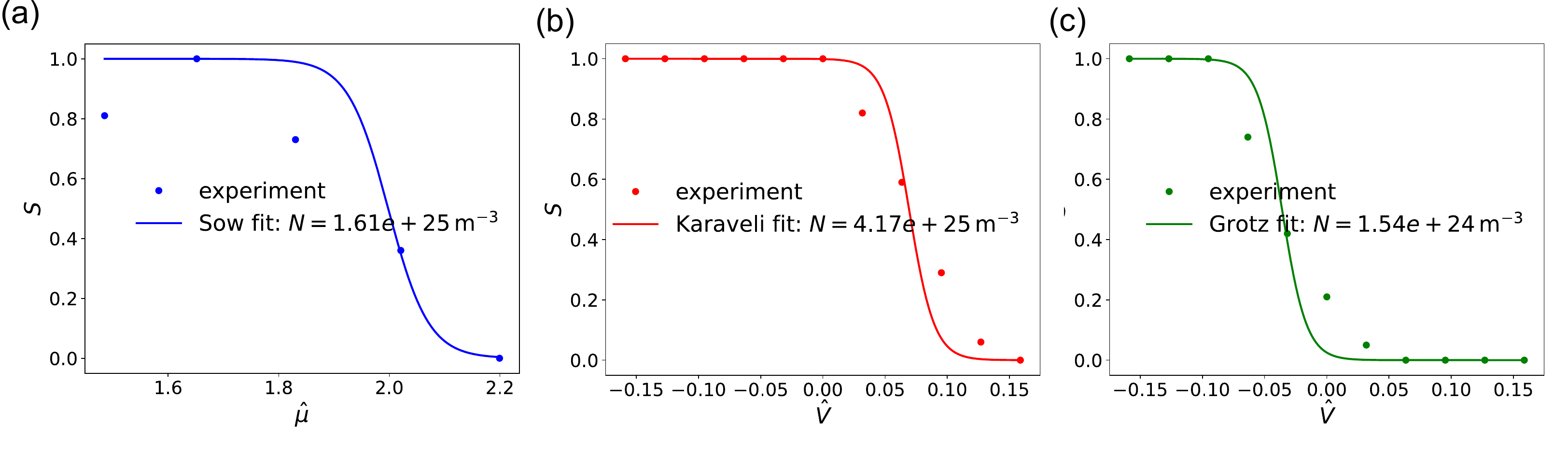}
\caption{Fitting of literature charge-state modulation data using the present electrostatic model. (a) pH-dependent fluorescence reported by Sow \textit{et al.}~\cite{sow_highthroughput_2020}. Voltage-dependent fluorescence reported by (b) Karaveli \textit{et al.}~\cite{karaveli_modulation_2016} and (c) Grotz \textit{et al.}~\cite{grotz_charge_2012}. Symbols denote manually digitized data normalized to the range $[0,1]$, and solid curves denote the model fits. The fitted values are $u_\mathrm{fit}=0.414\pm0.018$, $0.237\pm0.003$, and $0.199\pm0.003$ for panels (a)--(c), respectively. All three fitted values lie in the electrostatically transparent regime.}
  \label{fig:fitting}
\end{figure}

The three datasets used for fitting are shown in Fig.~\ref{fig:fitting}(a)--(c): (a) the pH-dependent fluorescence response reported by Sow \textit{et al.}~\cite{sow_highthroughput_2020}, and the voltage-dependent fluorescence responses reported by (b) Karaveli \textit{et al.}~\cite{karaveli_modulation_2016} and (c) Grotz \textit{et al.}~\cite{grotz_charge_2012}. All data were manually digitized and independently normalized to \([0,1]\) prior to fitting. For the Grotz \textit{et al.} dataset, the voltage sign was converted to match the sign convention for $V_\mathrm{b}$ used in the present model. All the parameters used in the fitting are summarized in Tables~\ref{tab:fitting_conditions} and \ref{tab:parameters}.

\begin{table*}[t]
\footnotesize
\caption{
Conditions, parameters, and fitting results for the previously reported charge-state responses.}
\label{tab:fitting_conditions}
\begin{ruledtabular}
\begin{tabular}{lccccccc}
Study & $I$ & pH & $V_\mathrm{b}$ & $x_0$ & Reference $N_\mathrm{D}$ range & \textsuperscript{g}$N_\mathrm{D, fit}^+$ & \textsuperscript{g}$u_\mathrm{fit}$ \\
& $(\mathrm{M})$ & & $(\mathrm{V})$ & $(10^{-9}~\mathrm{m})$ & $(10^{24}~\mathrm{m}^{-3})$ & $(10^{24}~\mathrm{m}^{-3})$ & \\

Sow \textit{et al.}~\cite{sow_highthroughput_2020}
& \textsuperscript{a}$1.00\times10^{-1}$
& $1,\ 4,\ 7.2,\ 10.2,\ 13$
& $0$
& \textsuperscript{d}$2.50$
& \textsuperscript{e}$8.80$--$70.4$
& $16.1\pm1.4$
& $0.414\pm0.018$
\\

Karaveli \textit{et al.}~\cite{karaveli_modulation_2016}
& \textsuperscript{b}$3.19\times10^{-2}$
& $7.7$
& $-0.50$--$0.50$
& \textsuperscript{d}$1.50$
& \textsuperscript{e}$8.80$--$70.4$
& $41.7 \pm 1.1$
& $0.237 \pm 0.003$ \\

Grotz \textit{et al.}~\cite{grotz_charge_2012}
& \textsuperscript{c}$7.00\times10^{-2}$
& $7.0$
& $-0.50$--$0.50$
& $7.00$
& \textsuperscript{f}$14.3$--$42.9$
& $1.54 \pm 0.04$
& $0.199 \pm 0.003$ \\

\end{tabular}
\end{ruledtabular}

\vspace{0.5em}
\noindent
\begin{minipage}{0.98\textwidth}
\raggedright
\footnotesize
\textsuperscript{a} The ionic strength was set to 0.1 M because the reported solution conditions do not uniquely determine ionic strength at pH 4, 7, and 10; and varying ionic strength over the plausible range of 0.001--1 M does not significantly affect the calculated signal.\\
\textsuperscript{b} Estimated from the reported liquid composition of 0.1 mM KCl, 10 mM Na$_2$HPO$_4$, and 1.8 mM KH$_2$PO$_4$.\\
\textsuperscript{c} Estimated from the reported liquid composition of 50 mM KCl and 10 mM potassium phosphate buffer.\\
\textsuperscript{d} For the nanodiamond datasets, the representative probe depth was estimated by assuming that the NV centers were uniformly distributed throughout a spherical particle of diameter $d$. The mean distance from an NV center at radial coordinate $r$ to the particle surface is $x_0=\frac{\int_{0}^{d/2}(d/2-r)4\pi r^2\,dr}{\int_{0}^{d/2}4\pi r^2\,dr}=d/8$. This estimate gives $x_0=2.50~\mathrm{nm}$ and $1.50~\mathrm{nm}$ for the nanodiamond diameters used by Sow \textit{et al.} and Karaveli \textit{et al.}, respectively.\\

\textsuperscript{e} Estimated from the reported nitrogen concentration range of 50--400 ppm for HPHT nanodiamonds~\cite{capelli_proximal_2022}.\\

\textsuperscript{f} For Grotz \textit{et al.}, the reference density was estimated from the reported nitrogen implantation dose of approximately $10^{13}~\mathrm{cm}^{-2}$. Assuming that the implanted nitrogen was uniformly distributed from the surface to the reported mean implantation depth of $7~\mathrm{nm}$, the average volume density is $N_{\mathrm{avg}}=(10^{13}~\mathrm{cm}^{-2})/(7~\mathrm{nm})=1.43\times10^{25}~\mathrm{m}^{-3}$.
Because the actual implantation profile is nonuniform and has a local maximum near the projected ion range, the reference interval was taken from this average density to three times its value, $1.43\times10^{25}$--$4.29\times10^{25}~\mathrm{m}^{-3}$.\\

\textsuperscript{g} The symmetric uncertainties in $N_\mathrm{D, fit}^+$ and $u_\mathrm{fit}$ denote rounded 95\% confidence intervals obtained from the fits.
\end{minipage}
\end{table*}

\subsection{Fitting results}

The fitted ionized-donor densities $N_\mathrm{D,fit}^{+}$ and the corresponding universal scaling factors $u_\mathrm{fit}$ are summarized in Table~\ref{tab:fitting_conditions}. The resulting $u_\mathrm{fit}$ values range from approximately 0.20 to 0.41, placing all three datasets in the electrostatically transparent regime. For all three datasets, the fitted ionized-donor densities lie within or below the corresponding reference nitrogen-density ranges, consistent with the interpretation of $N_\mathrm{D,fit}^{+}$ as an effective ionized-donor density rather than the total or implanted nitrogen concentration.

\section{Parameter values used in calculations}
All numerical calculations in this study were performed using the parameters summarized in Table~\ref{tab:parameters}, unless otherwise specified.

\section{Applicability and extensions of the model}
The universal scaling derived above follows from the idealized electrostatic assumptions stated in Sec.~S1. Applying the model quantitatively to a specific experiment may require additional material and operational effects. These corrections change the mapping between experimental parameters and the effective electrical lengths, but they do not by themselves alter the logic of expressing information transfer in terms of electrostatic propagation and separation scales.

First, real solid surfaces can carry chemically active functional groups whose charge state depends on the local solution environment. Such groups form an interfacial surface-charge layer and modify the boundary condition for the displacement field. Within the present framework, this effect can be incorporated by adding a two-dimensional surface charge density at $x=0$ and solving the modified electrostatic boundary-value problem.

Second, finite solid size becomes important when the solid dimension is comparable to the depletion width, particularly for nanoparticles that can become fully depleted. In this regime, the semi-infinite planar solution used in Secs.~S1--S2 is no longer sufficient, and an additional electrostatic boundary condition can be imposed at the opposite surface.

Third, the effective ionized-dopant density may differ substantially from the nominal total dopant or nitrogen concentration and is often not known independently. It can therefore be treated as an effective parameter constrained by independent measurements or by fitting the electrostatic response, as done in Sec.~S5. This uncertainty affects the inferred value of $L_\mathrm{s}$ and hence the quantitative mapping from a particular material to $u$.

Fourth, the ideal Fisher information assumes unit-fidelity binary readout. Experimental readout errors reduce the accessible information below $\mathcal{I}^*$, so improved charge-state discrimination and readout protocols are required to approach the theoretical limit~\cite{alsid_photoluminescence_2019,zhao_suppression_2012,audecraik_microwaveassisted_2020,chakraborty_magneticfieldassisted_2022,thalassinos_robust_2025}. Similarly, an externally applied bias need not drop exclusively across the solid--liquid electrostatic structure considered here; part of the voltage can be distributed across the electrolyte, contacts, interfaces, or other circuit elements. A platform-specific voltage-partition model is therefore required when converting the externally applied voltage into $V_\mathrm{b}$ used in the theory.

Finally, beyond the Debye--H\"uckel regime specified in Sec.~S1, nonlinear screening, ion-specific interactions, and concentrated-solution effects can modify the liquid-side potential. In such cases, the electrolyte should be described by the nonlinear Poisson--Boltzmann equation or a more detailed interfacial model. These extensions define the range over which the present closed-form scaling can be applied quantitatively.

\begin{table*}[t]
\caption{Calculation settings.}
\label{tab:parameters}
\centering
\begin{tabular}{lll}
\hline\hline
Symbol & Description & Value \\
\hline
\multicolumn{3}{l}{\textit{Physical constants}} \\
$e$ & Elementary charge & $1.602176634\times10^{-19}~\mathrm{C}$ \\
$k_\mathrm{B}$ & Boltzmann constant & $1.380649\times10^{-23}~\mathrm{J\,K^{-1}}$ \\
$T$ & Temperature & $300.0~\mathrm{K}$ \\
$N_\mathrm{A}$ & Avogadro constant & $6.02214076\times10^{23}~\mathrm{mol^{-1}}$ \\
$\varepsilon_0$ & Vacuum permittivity & $8.8541878128\times10^{-12}~\mathrm{F\,m^{-1}}$ \\
\hline
\multicolumn{3}{l}{\textit{Liquid}} \\
$\varepsilon_\mathrm{e}$ & Liquid permittivity & $80.0\varepsilon_0$ \\
$E_\mathrm{SHE}^{\mathrm{abs}}$ & Absolute potential of the standard hydrogen electrode & $4.44~\mathrm{V}$ \\
$E_{\mathrm{O_2/H_2O}}^\circ$ & Standard electrode potential for the $\mathrm{O_2/H_2O}$ redox couple & $1.229~\mathrm{V}$ \\
$n$ & Number of electrons involved in the $\mathrm{O_2/H_2O}$ redox couple & $4$ \\
$p_{\mathrm{O_2}}$ & Oxygen partial pressure & $0.21$ \\
$I$ & Ionic strength & $1~\mathrm{mol\,m^{-3}}$ \\
$\mathrm{pH}$ & pH of the liquid & $7$ \\
\hline
\multicolumn{3}{l}{\textit{Solid}} \\
$\varepsilon_\mathrm{s}$ & Solid permittivity & $5.6\varepsilon_0$ \\
$N_\mathrm{D}^{+}$ & Ionized dopant density & $1.0\times10^{24}~\mathrm{m^{-3}}$ \\
$\chi$ & Electron affinity & $-1.1~\mathrm{eV}$ \\
$E_\mathrm{C}$ & Electron energy at the conduction-band minimum & $-\chi$ \\
$E_\mathrm{G}$ & Band gap & $5.5~\mathrm{eV}$ \\
$E_\mathrm{V}$ & Electron energy at the valence-band maximum & $E_\mathrm{C}-E_\mathrm{G}$ \\
$E_\mathrm{F0}$ & Reference Fermi level & $E_\mathrm{C}-1.7~\mathrm{eV}$ \\
$E_\mathrm{T0}$ & Charge-transition energy & $E_\mathrm{V}+2.8~\mathrm{eV}$ \\
$V_\mathrm{b}$ & Bias voltage & $0~\mathrm{V}$ \\
$x_0$ & Probe depth & $4.0\times10^{-9}~\mathrm{m}$ \\
\hline\hline
\end{tabular}
\end{table*}

\end{document}